\documentclass[a4paper]{spie}  %>>> use this instead for A4 paper
\usepackage{graphicx}
\usepackage{tabularx}
\usepackage{makecell}
\usepackage{array}
\usepackage{subcaption}
\usepackage{caption}
\usepackage{amsmath,amsfonts,amssymb}
\usepackage{graphicx}
\usepackage[colorlinks=true, allcolors=blue]{hyperref}

\title{Self-supervised Registration and Segmentation of the Ossicles with A Single Ground Truth Label}

\author[a]{Yike Zhang}
\author[a,b]{Jack H. Noble}
\affil[a]{Dept. of Computer Science, Vanderbilt University}
\affil[b]{Dept. of Electrical and Computer Engineering, Vanderbilt University}

\authorinfo{Further author information: (Send correspondence to Yike Zhang)\\Yike Zhang: E-mail: yike.zhang@vanderbilt.edu, Telephone: 1 210 449 0900\\ Jack Noble: E-mail: jack.noble@vanderbilt.edu, Telephone: 1 615 875 5539}

\begin{document} 
\maketitle

\begin{abstract}
% Introduce topic
AI-assisted surgeries have drawn the attention of the medical image research community due to their real-world impact on improving surgery success rates. For image-guided surgeries, such as Cochlear Implants (CIs), accurate object segmentation can provide useful information for surgeons before an operation. 
% Specific task and problem
Recently published image segmentation methods that leverage machine learning usually rely on a large number of manually predefined ground truth labels. However, it is a laborious and time-consuming task to prepare the dataset. 
% Methodology
This paper presents a novel technique using a self-supervised 3D-UNet that produces a dense deformation field between an atlas and a target image that can be used for atlas-based segmentation of the ossicles.
% Results
Our results show that our method outperforms traditional image segmentation methods and generates a more accurate boundary around the ossicles based on Dice similarity coefficient and point-to-point error comparison. The mean Dice coefficient is improved by 8.51\% with our proposed method.

\end{abstract}


\section{INTRODUCTION}
\label{sec:intro}  % \label{} allows reference to this section
% Surgery Background
Surgical procedures require accurate medical segmentation tools to assist surgeons in decision-making and planning. The ossicles, consisting of the malleus, incus, and stapes, are the smallest bones in the body. They serve as a bridge between the tympanic membrane and the inner ear. When determining the best cochlear implant(CI) insertion vector, they represent an important landmark to assist surgeons. They can also be used as an intra-operative landmark for following a pre-operatively planned insertion vector\cite{rf}. Thus, accurate segmentation of the ossicles in a pre-operatively acquired CT scan is needed for effective pre-operative planning and intra-operative guidance. 

% Background literature
Noble et al.\cite{jh1} propose an atlas-based segmentation approach that uses a traditional nonrigid registration method called the adaptive bases algorithm to generate a deformation field that registers the target image and an atlas where the ossicles are accurately segmented \cite{gua}. The deformation field is used to project the ossicles segmentation from the atlas to the target, thereby segmenting the ossicles. The authors report mean surface distance errors for this approach of approximately 0.4mm, indicating results that are consistently close, but not sub-voxel accurate. Another recent work also proposed a traditional atlas-based approach, resulting in mean surface distance errors between 0.1 and 0.157 mm for the ossicles \cite{andy}. Wang et al.\cite{wang2021} propose a deep neural network solution for segmenting the ossicles and other structures inside the ear. The network produces a probability map for the ossicles at the same resolution as the input image. They report the method results in mean surface distance errors of 0.107 mm on a dataset of normal anatomy. However, fine-scale subregions of the ossicles, such as the stirrup shape of the stapes, cannot be reconstructed by this approach due to partial volume effects arising from the coarse resolution of the image relative to the size of the structure. Further, as it predicts a probability map, rather than a coordinate transformation, it is not possible to localize various sub-regions of the structure being segmented. Obtaining a coordinate transformation is useful for surgery planning, where it would be used to measure distances between different subregions of the structure relevant for the surgery plan. For example, providing measurements of the distance and orientation between a pre-operatively planned CI insertion trajectory and the stapes footplate or the incus-stapes joint can assist surgeons in realizing the pre-operatively planned trajectory. 

% Discuss Unet network
U-Net\cite{unet} is a widely adopted deep learning architecture that outperforms prior Convolutional Neural Network (CNN\cite{cnn}) based segmentation methods. It aims for precise image segmentation and outputs superior results when testing on medical image datasets. Our proposed deep learning network architecture is based on the original U-Net structure and it achieves satisfying results. 

Non-rigid deformable deep learning models has drawn the attention of the medical imaging research community in recent years because of their data efficiency and expressivity. As a promising deep learning approach, VoxelMorph\cite{bal} is an unsupervised pairwise 3D medical image registration framework. The model takes a pair of images (moving image and fixed image) as input, and it directly generates a smooth dense deformation field using a spatial transform layer that registers the moving image to the fixed image. The learning-based algorithm is fast and suitable for medical image analysis and processing.

% Introduction of our method
In this work, we propose a deep learning network based model to perform atlas-based segmentation of the ossicles. Inspired by VoxelMorph, we propose a self-supervised U-Net-based network that produces a dense deformation field that registers the input target image to an atlas image. The highlight of our approach is that the main task of the network is to segment a specific structure, and thus we propose a region-based level-set inspired loss function that rewards deformation fields which optimize the segmentation to align with the structure's boundaries.

\section{METHODS}
\label{sec:methods}
The proposed 3D U-Net-based model is used to perform registration between the atlas image $m$ and a patient CT scan $p$, thus, obtaining a deformation field $\phi$ mapping coordinates from $m$ to $p$. Fig. 1 presents the overall architecture. It consists of eight encoder layers, three max-pooling layers, and ten decoder layers. During the experiments, our input $p$ size is 1 × 64 × 76 × 44. This is different from VoxelMorph since we focus on registering CT scans to a single atlas scan. We reduce the network input architecture to include only the target image as the input rather than adding the atlas image to a separate channel. The red outlined encoder blocks mean that we apply Volumetric Batch Normalization before forwarding it to the next layer. Our dataset includes 655 patient CT volumes, which are randomly split into proportions of 0.7, 0.2, and 0.1 for training, validating, and testing, respectively. 
\begin{figure}[ht]
\centering
\includegraphics[width=\textwidth]{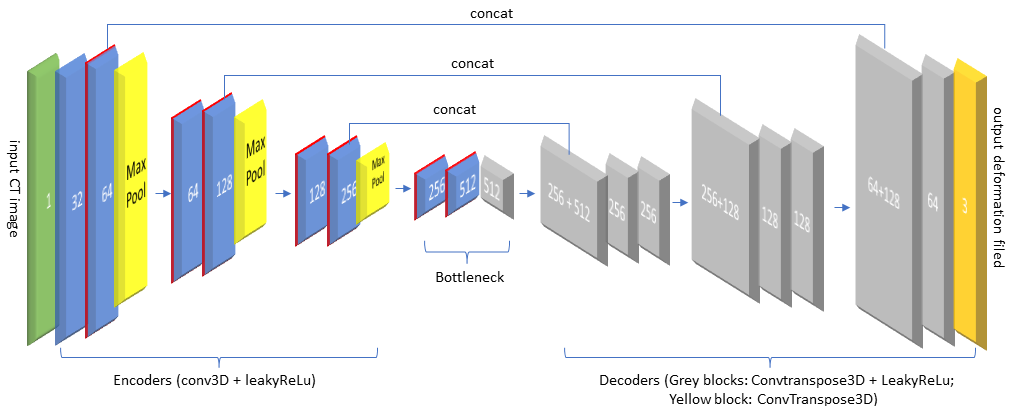}
\vspace{0.2cm}
\caption{Proposed 3D-UNet structure.}
\end{figure}
\label{fig: unet3d_structure}

$\phi^{-1}(p)$ denotes a patient image volume $p$ projected through the nonrigid deformation field $\phi$ for registration with the atlas. There is one manually defined atlas surface mesh $\alpha$ defining the ossicles segmentation, and a corresponding atlas mask $\beta$ in the dataset. $\phi$($\alpha$) denotes the atlas ossicles surface projected forward through the deformation to the target image to produce an atlas-based segmentation. At the end of the training, the refined deformation field $\phi$ is applied on the atlas mesh $\alpha$ to obtain $\phi(\alpha)$ and thus resulting in highly accurate segmentation. For the loss functions, we use the cross-correlation loss $L_{cc}$ as the self-supervision term to optimize the overall intensity mapping between the atlas image $m$ and the deformed target image $\phi^{-1}(p)$. The $L_{cc}$ between $m$ and $\phi^{-1}(p)$ is defined in Eq. \ref{Eq: eq1}, where $N$ is noted as the total number of pixels.
\begin{equation}
    L_{cc}(m, \phi^{-1}(p)) = \frac{\sum_{i=1}^{N}\left [\left( m_{i} - \overline{m} \right) \left( \phi^{-1}(p)_{i} - \overline{\phi^{-1}(p)} \right) \right ]}{\sqrt{\sum_{i=1}^{N}\left(  m_{i} - \overline{m}  \right)^2 \sum_{i=1}^{N}\left(  \phi^{-1}(p)_{i} - \overline{\phi^{-1}(p)}  \right)^2}}
    \label{Eq: eq1}
\end{equation} The gradient 3D loss $L_{gd}$ in Eq. \ref{Eq: eq2} is a regularization function applied to ensure the deformation field $\phi$ output by the network is smooth and continuous.
\begin{equation}
    L_{gd}(\phi) = \sum_{i=1}^{N}\left\| \nabla \left( \phi_{i} \right) \right\|^2
\label{Eq: eq2}
\end{equation} 
In this equation, we define $\frac{\partial{\phi}}{\partial{x}} \approx \phi(i_x+1, i_y, i_z) - \phi(i_x, i_y, i_z)$, and the same rule applies to $\frac{\partial{\phi}}{\partial{y}}$ and $\frac{\partial{\phi}}{\partial{z}}$. Finally, we utilize a custom level-set method\cite{levelset} inspired loss $L_{ls}$ to refine the intensity mapping in the vicinity of the target structure, thus obtaining a more accurate deformation field near the structure and a more accurate atlas-based segmentation boundary. $L_{ls}$ takes three inputs: $\phi^{-1}(p)$, $\beta$, and $\mu$ --- which is a dilated version of $\beta$. $\beta$ is dilated by a spherical structuring element with radius equal to 3.0 voxel size. The goal is to maximize the difference between the target image intensities of $\phi^{-1}(p)$ falling within foreground $\beta \cdot \mu$, which we term $L_{lsf}$, and the target image intensities of $\phi^{-1}(p)$ falling within background $(1 - \beta) \cdot \mu$, which we term $L_{lsb}$. We limit the range of the background region considered in the loss function to the region included in the foreground of $\mu$, so that other surrounding bone structures that are not immediately adjacent to the ossicles are not included.
\begin{align}
    L_{ls}(\phi^{-1}(p), \beta, \mu) &= -(L_{lsf} - L_{lsb}) \\
    \\
    &= -\frac{\sum \phi^{-1}(p) \cdot \mu \cdot (\beta - (1-\beta))}{\sum\mu} \\
    \\
    &= -\frac{\sum \phi^{-1}(p) \cdot \mu \cdot (2\beta - 1))}{\sum\mu}
\end{align}
The overall loss function can be summarize as:
\begin{equation}
    Loss = \lambda_{cc} \cdot L_{cc}(m, p(\phi)) + \lambda_{gd} \cdot L_{gd}(\phi) + \lambda_{ls} \cdot L_{ls}(p(\phi), \beta, \mu)
\end{equation} Where $\lambda_{cc}$, $\lambda_{gd}$ and $\lambda_{ls}$ represent different weights for the three loss terms. In our experiments, we use 0.1 for $\lambda_{cc}$, 0.85 for $\lambda_{gd}$ and 0.05 for $\lambda_{ls}$ found through manual hyper-parameters tuning. Fig. \ref{fig: no_levelset_loss} is included to compare the performance differences when training without using $\lambda_{ls}$.

\section{RESULTS}
\label{sec:results}
\begin{figure}[ht]

\newcommand{\ImgRatio}{0.245}
\newcommand{\ImgWidth}{1}
\begin{center}
    % first row
\begin{subfigure}{\ImgRatio\textwidth}
  \centering
  % include first image
  \includegraphics[width=\ImgWidth\linewidth]{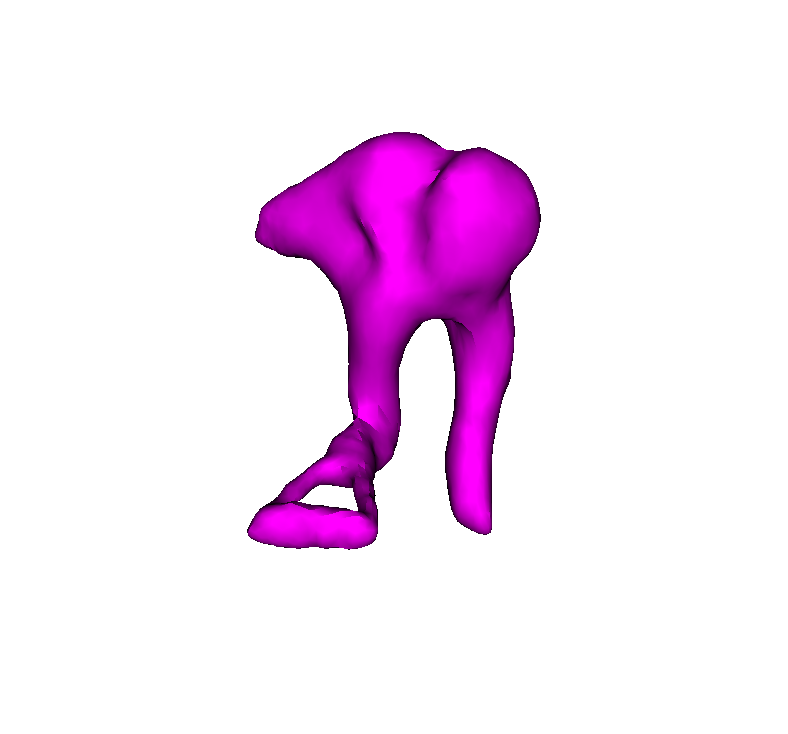}
  \caption{3D view}
\end{subfigure}
\begin{subfigure}{\ImgRatio\textwidth}
  \centering
  % include second image
  \includegraphics[width=\ImgWidth\linewidth]{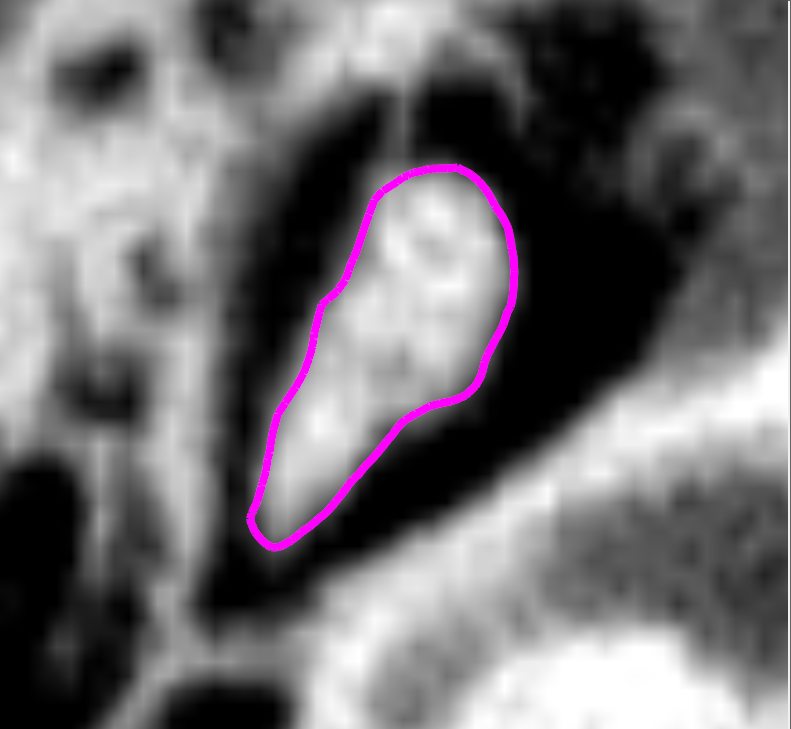}
  \caption{Axial View in CT}
\end{subfigure}
\begin{subfigure}{\ImgRatio\textwidth}
  \centering
  % include second image
  \includegraphics[width=\ImgWidth\linewidth]{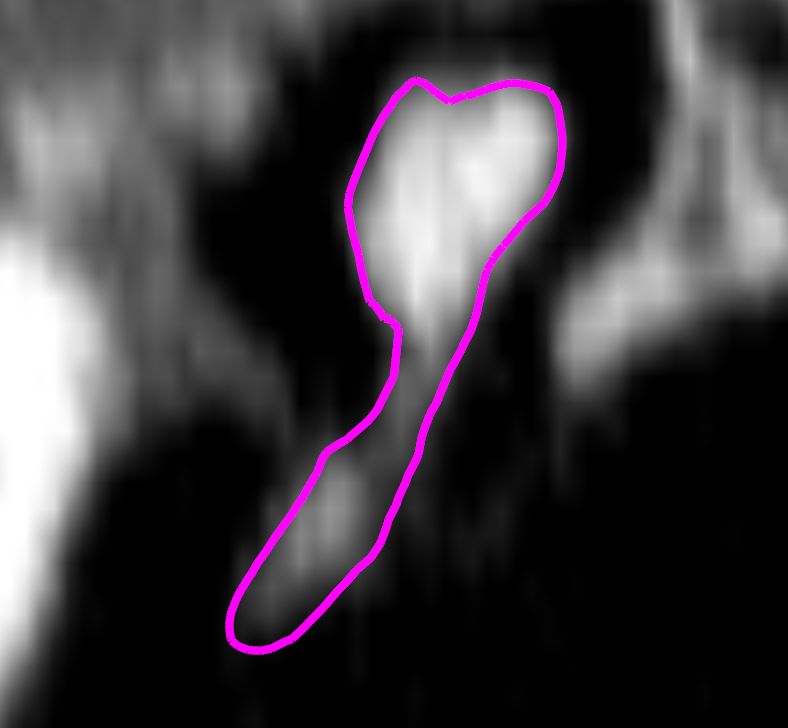}
  \caption{Coronal View in CT}
\end{subfigure}
\begin{subfigure}{\ImgRatio\textwidth}
  \centering
  % include second image
  \includegraphics[width=\ImgWidth\linewidth]{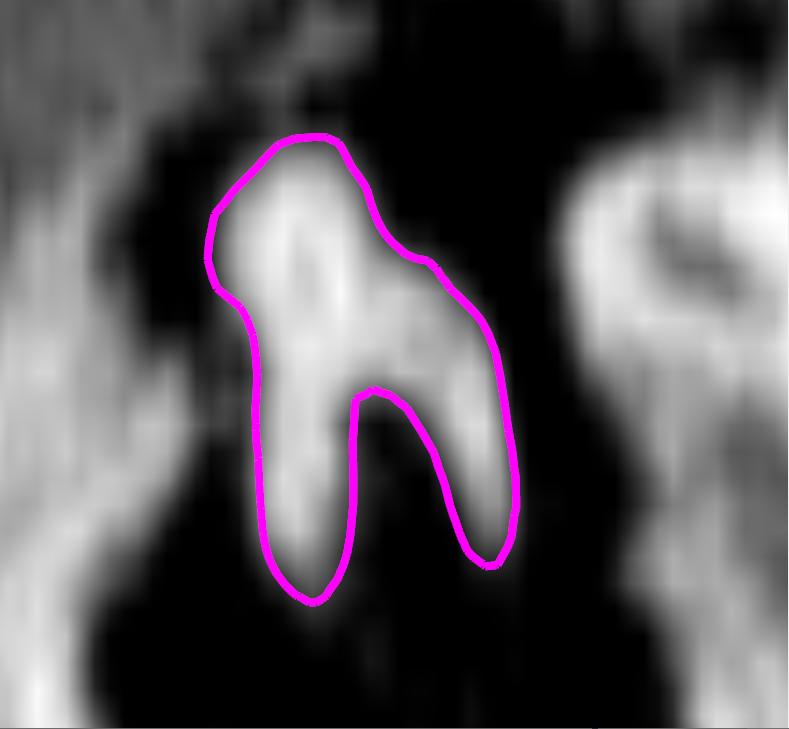}
  \caption{Sagittal View in CT}
\end{subfigure}

\end{center}
\caption{An overview of the output ossicles.}
\label{fig: overview}
\end{figure}

\begin{figure}[ht]

\newcommand{\ImgRatio}{0.24}
\newcommand{\ImgWidth}{0.95}

\centering
% \captionsetup[subfigure]{labelformat=empty}

\begin{subfigure}{\ImgRatio\textwidth}
  \centering
  % include first image
  \includegraphics[width=\ImgWidth\linewidth]{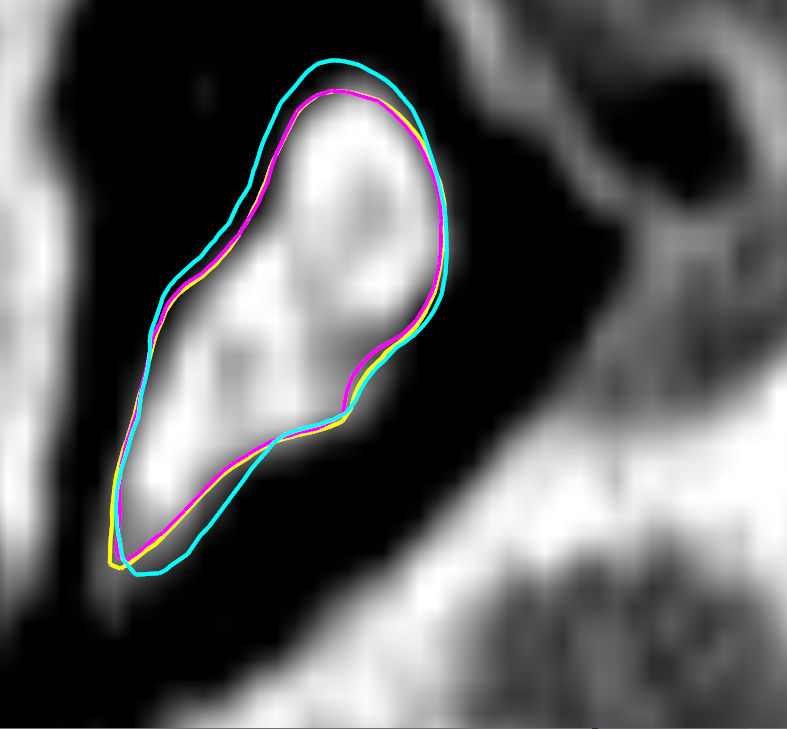}
  \caption{}

\end{subfigure}
\begin{subfigure}{\ImgRatio\textwidth}
  \centering
  % include second image
  \includegraphics[width=\ImgWidth\linewidth]{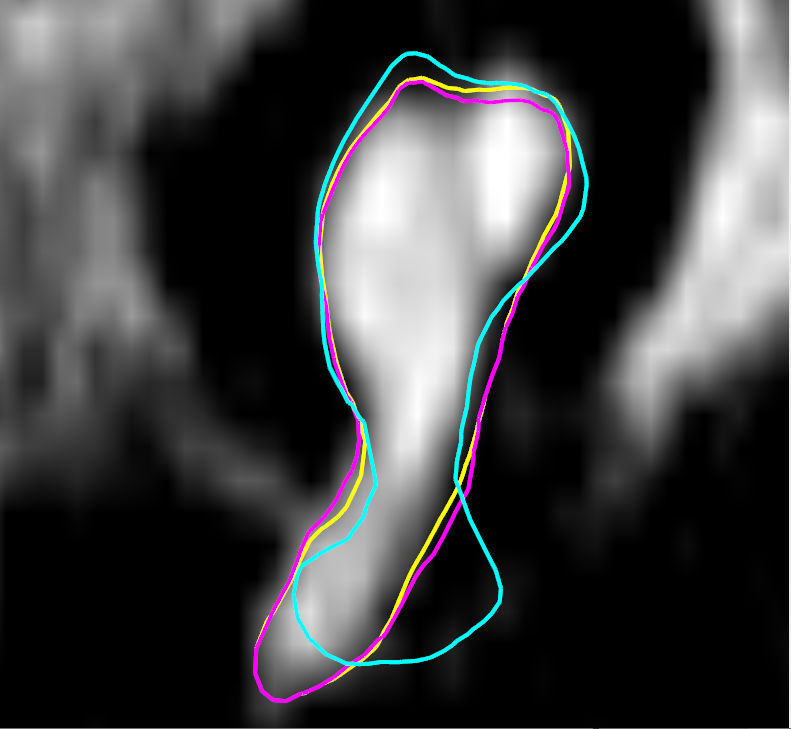}
  \caption{}

\end{subfigure}
\begin{subfigure}{\ImgRatio\textwidth}
  \centering
  % include second image
  \includegraphics[width=\ImgWidth\linewidth]{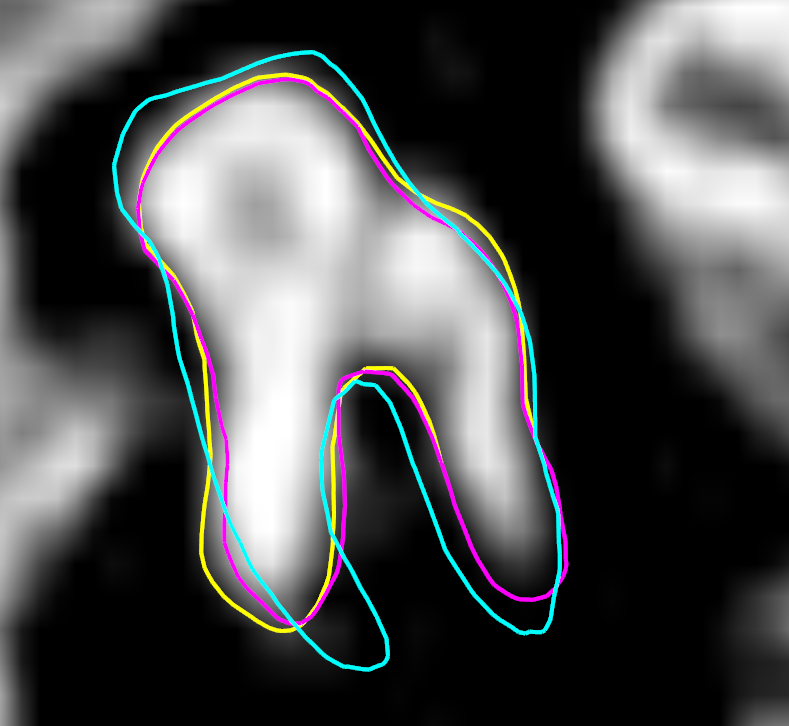}
  \caption{}

\end{subfigure}
\begin{subfigure}{\ImgRatio\textwidth}
  \centering
  % include second image
  \includegraphics[width=\ImgWidth\linewidth]{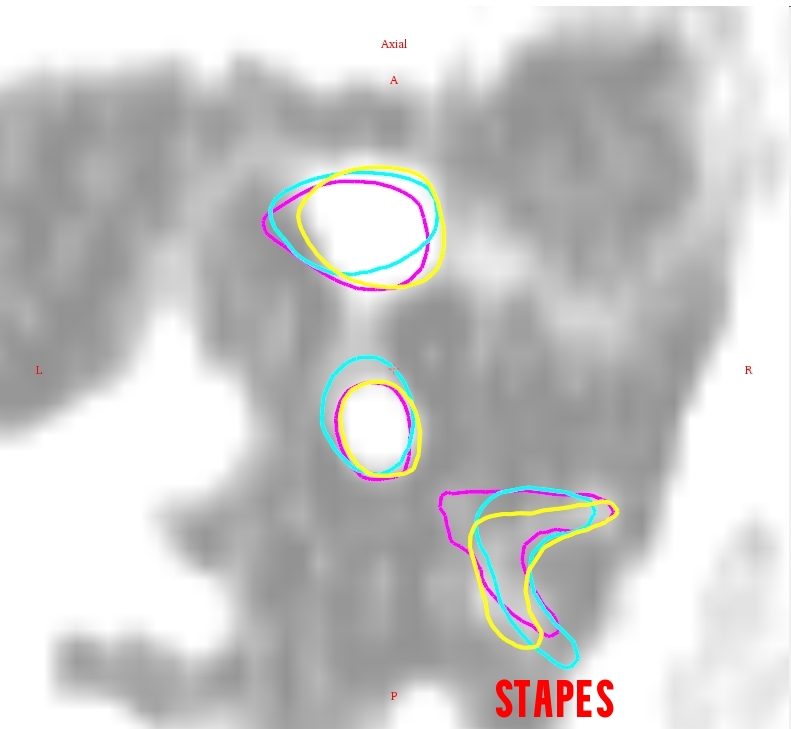}
  \caption{}
  \label{fig:stapes_view}

\end{subfigure}

\begin{subfigure}{\ImgRatio\textwidth}
  \centering
  % include first image
  \includegraphics[width=\ImgWidth\linewidth]{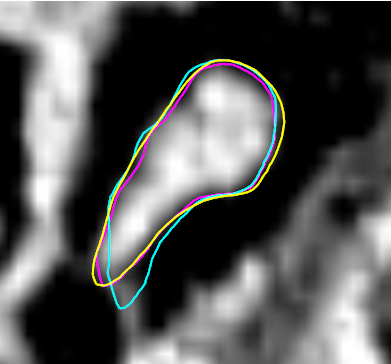}
  \caption{}

\end{subfigure}
\begin{subfigure}{\ImgRatio\textwidth}
  \centering
  % include second image
  \includegraphics[width=\ImgWidth\linewidth]{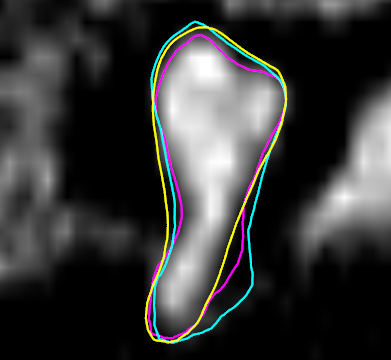}
  \caption{}

\end{subfigure}
\begin{subfigure}{\ImgRatio\textwidth}
  \centering
  % include second image
  \includegraphics[width=\ImgWidth\linewidth]{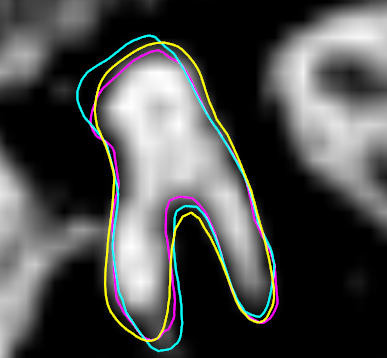}
  \caption{}

\end{subfigure}
\begin{subfigure}{\ImgRatio\textwidth}
  \centering
  % include second image
  \includegraphics[width=\ImgWidth\linewidth]{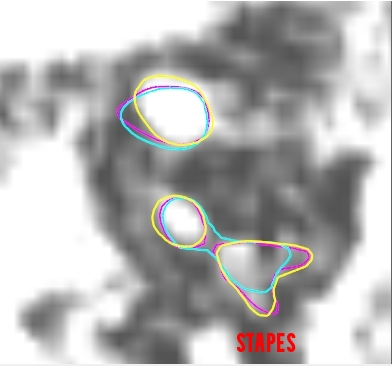}
  \caption{}

\end{subfigure}

\vspace{.2cm}
\caption{First three columns are comparing three different views (Axial, Coronal, and Sagittal view) of two outputs. The last column is the comparison of the stapes segmentation.}
\end{figure}

An output example obtained from our trained neural network is presented in Fig. \ref{fig: overview}. As can be seen, the network creates an atlas-based segmentation that well agrees with the visually apparent boundaries of the ossicles. We compare the sagittal, coronal, and axial views of random selected outputs obtained from the test dataset in Fig. \ref{fig:compare} (a)-(c). Magenta outlines the network output results, aqua represents the adaptive bases algorithm results, and yellow indicates the manually segmented ground truth labels. Fig. \ref{fig:stapes_view} is the comparison in an axial slice that includes the challenging stapes portion of the ossicles, where a brighter intensity window is used to make the stapes visible. Network outputs more closely resemble the manually segmented ground truth labels than adaptive bases algorithm outputs. 
\label{fig:compare}

\begin{figure}[ht]
  % Fixed length
  \centering
  \subcaptionbox{Dice Score Comparison \label{fig:box_plots1}}{\includegraphics[width=3.15in, trim=0cm 2cm 2cm 0cm,clip]{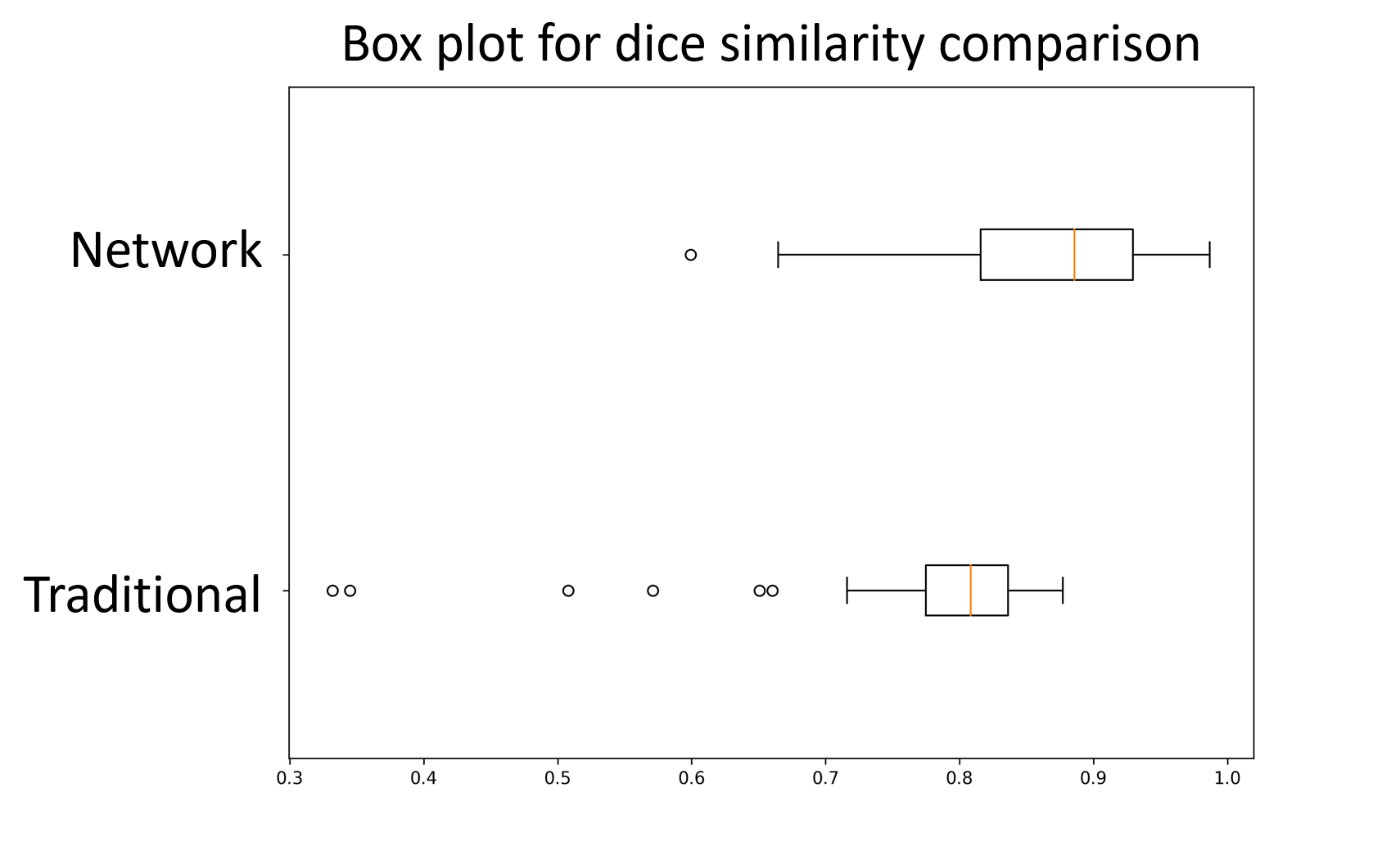}} \hspace{3em}
  \centering
  \subcaptionbox{Point-to-point Error Comparison \label{fig:box_plots2}}{\includegraphics[width=2.9in, trim=1cm 0cm 0cm 0cm,clip]{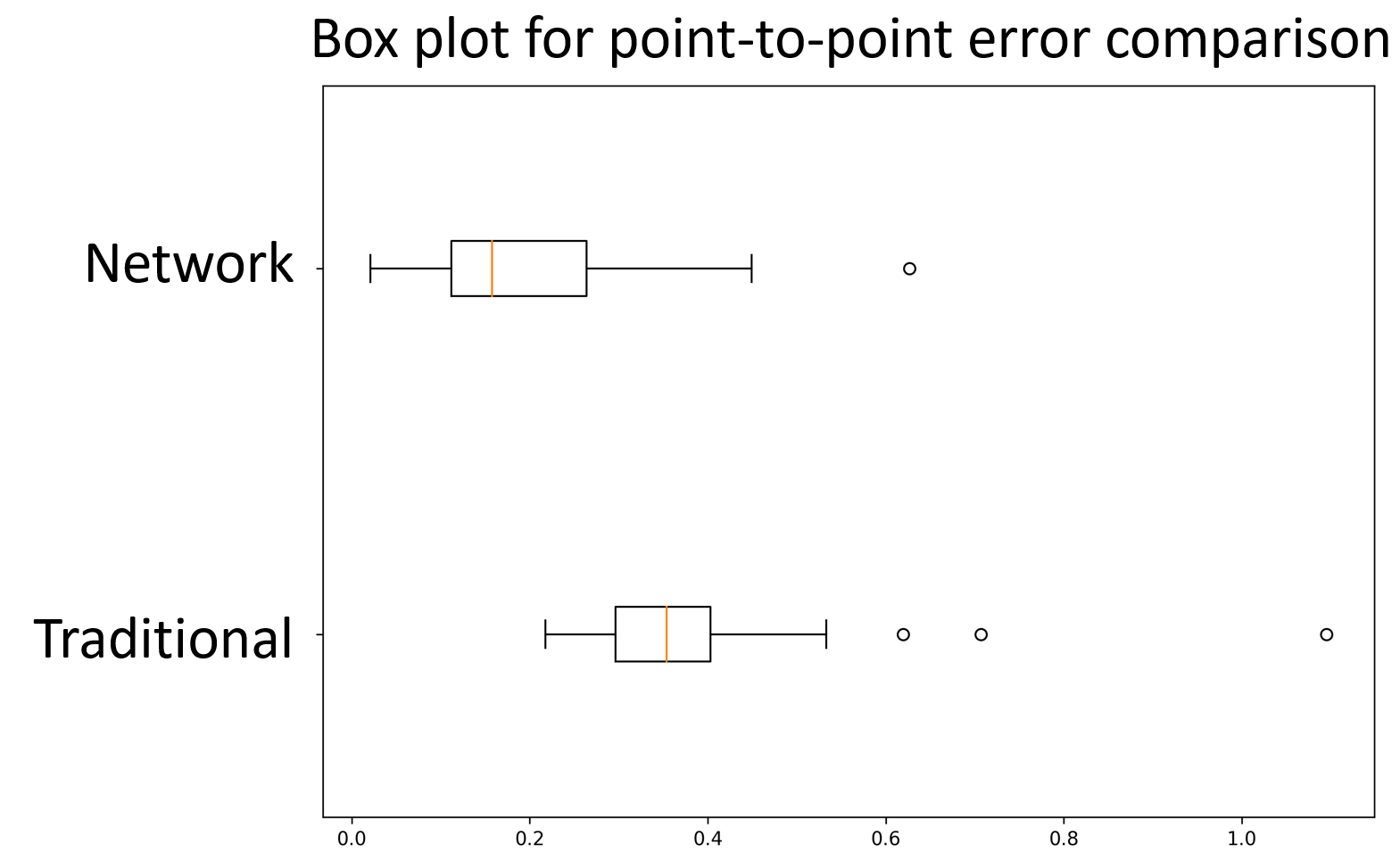}}
\vspace{2mm}
\caption{Box plots for 64 samples in the testing dataset.}
\label{fig:box_plots}
\end{figure}

To quantitatively compare the similarity using dice score between the proposed method, traditional method (adaptive bases algorithm), and the ground truth labels, we provide a box plot in Fig. \ref{fig:box_plots1} for 64 samples in the testing dataset. The mean dice score for the network outputs is 0.8665 and for the adaptive bases algorithm is 0.7814. The paired differences comparison p-value from Wilcoxon signed rank test is $p=8.2975e-10$. Fig. \ref{fig:box_plots2} plot is for point-to-point error comparison based on the Euclidean distance between homologous mesh points in millimeters(mm). It is shown that the meshes generated from the network have smaller errors than the meshes obtained from the adaptive bases algorithm. The mean point-to-point error for network outputs is 0.1914mm and for the adaptive bases algorithm outputs is 0.3759mm. The low point-to-point error achieved by the network indicates high accuracy in localizing landmarks in the different ossicles subregions. Further, the Mean Surface Distance (MSD) error between the network outputs and ground truth labels is 0.0897mm while the MSD error between adaptive bases algorithm outputs and ground truth labels is 0.1826mm. %Since we don't have ground truth samples in our original dataset, we generated the ground truth labels directly from the network output since the ossicles boundaries are already visually apparent. There may be bias exists in such generating fashion, and we will include more analysis in the future work. 

% \begin{figure}[ht]
% % \captionsetup[subfigure]{labelformat=empty}

% % first row
% \begin{subfigure}{.5\textwidth}
%   \centering
%   % include first image
%   \includegraphics[width=.8\linewidth]{figures/box_plot1.png}
%   \caption{Performance Comparison}
% \end{subfigure}
% \hspace{-0.2cm}
% \begin{subfigure}{.6\textwidth}
%   \centering
%   % include second image
%   \includegraphics[width=.65\linewidth]{figures/box_plot2.png}
%   \caption{Pair differences comparison}

% \end{subfigure}
% \vspace{0.1cm}

% \caption{Box plots for 22 samples in the testing dataset.}
% \label{fig:box_plots}
% \end{figure}
\vspace{-3mm}
\vspace{2mm}

% \begin{figure}[ht]

% \newcommand{\ImgRatio}{0.4}
% \newcommand{\ImgWidth}{1.1}
% \newcommand{\ImgWidt}{1.1}

% \centering
% % \captionsetup[subfigure]{labelformat=empty}

% \begin{subfigure}{\ImgRatio\textwidth}
%   \centering
%   \captionsetup[subfigure]{justification=centering}
%   % include first image
%   \includegraphics[width=\ImgWidth\linewidth, trim=0cm 2cm 2cm 0cm,clip]{figures/box_plot1.png}
%   \caption{Dice Score Comparison}
%   \label{fig:box_plots1}
  
% \hspace{2cm}
% \end{subfigure}
% \begin{subfigure}{\ImgRatio\textwidth}
%   \centering
%   \captionsetup[subfigure]{justification=centering}
%   % include second image
%   \includegraphics[width=\ImgWidt\linewidth, trim=1cm 0cm 0cm 0cm,clip]{figures/box_plot2.png}
%   \caption{Point-to-point Error Comparison}
%   \label{fig:box_plots2}

% \end{subfigure}
% \vspace{.2cm}
% \caption{Box plots for 64 samples in the testing dataset.}
% \label{fig:box_plots}
% \end{figure}

\begin{figure}[ht]

\newcommand{\ImgRatio}{0.32}
\newcommand{\ImgWidth}{0.95}

\centering
% \captionsetup[subfigure]{labelformat=empty}

\begin{subfigure}{\ImgRatio\textwidth}
  \centering
  % include first image
  \includegraphics[width=\ImgWidth\linewidth]{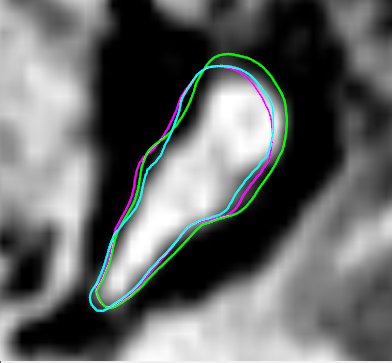}
  \caption{}

\end{subfigure}
\begin{subfigure}{\ImgRatio\textwidth}
  \centering
  % include second image
  \includegraphics[width=\ImgWidth\linewidth]{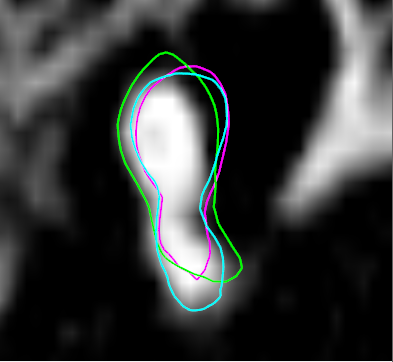}
  \caption{}

\end{subfigure}
\begin{subfigure}{\ImgRatio\textwidth}
  \centering
  % include second image
  \includegraphics[width=\ImgWidth\linewidth]{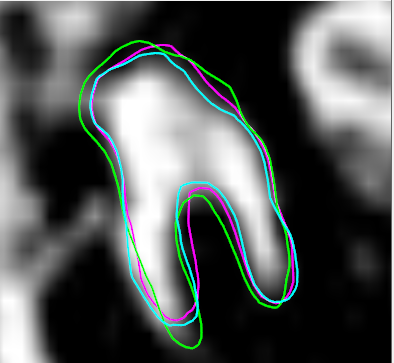}
  \caption{}

\end{subfigure}

\begin{subfigure}{\ImgRatio\textwidth}
  \centering
  % include second image
  \includegraphics[width=\ImgWidth\linewidth]{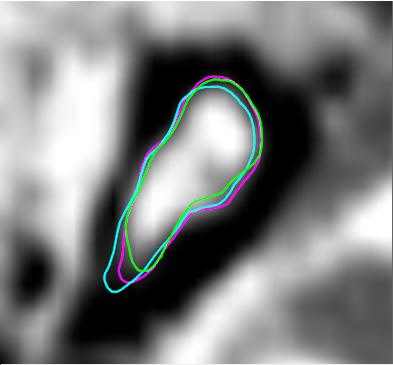}
  \caption{}

\end{subfigure}
\begin{subfigure}{\ImgRatio\textwidth}
  \centering
  % include second image
  \includegraphics[width=\ImgWidth\linewidth]{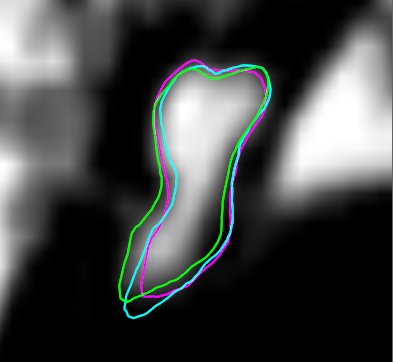}
  \caption{}

\end{subfigure}
\begin{subfigure}{\ImgRatio\textwidth}
  \centering
  % include second image
  \includegraphics[width=\ImgWidth\linewidth]{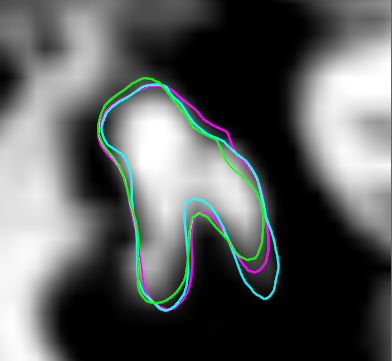}
  \caption{}

\end{subfigure}

\vspace{.2cm}
\caption{Performance comparison. Blue contour shows the outputs from network trained with using the level-set loss. Magenta contour shows the outputs from network trained without using level-set inspired loss, and green contour shows the output produced by adaptive bases algorithm.}
\label{fig: no_levelset_loss}
\end{figure}

Fig. \ref{fig: no_levelset_loss} demonstrates the performance differences with random selected samples. The outputs highlighted in magenta contour are from trained network without using the level-set inspired loss $\lambda_{ls}$ while the blue contour represents the results when training network with $\lambda_{ls}$. The green contour represents results that output by adaptive bases algorithm. The figure below show that when training with $\lambda_{ls}$, the overall performance is generally better than training without using $\lambda_{ls}$.

\section{CONCLUSION}
\label{sec:conclusion} 
Our paper presents a novel atlas-based segmentation deep learning network that only utilizes a single ground truth with many unlabeled images to achieve promising results through self-supervised learning. In contrast to standard mask-based segmentation methods, atlas-based segmentation methods provide a coordinate mapping that can be used to easily identify sub-regions and local landmarks of structures of interest. Furthermore, we discussed that it is possible to use only one ground truth to train a self-supervised 3D UNet to achieve excellent results. Thus, laborious and expensive manual medical image labeling can be avoided. The proposed work is highly adaptable and has the potential to solve many similar medical segmentation tasks. Mean surface errors achieved by our method are lower than other existing ossicles segmentation methods.

One limitation of this work is in the construction of the ground truth. In order to create ground truth ossicles segmentation surfaces that have one-to-one vertex correspondence with the atlas, the automatic method was first used to initialize the ground truth surface. These initializations were then manually edited in designed software to correct errors. Due to this initialization method, our quantitative results are likely biased towards the automated method. Another limitation is the lack of investigation of different deep learning architectures. We use a traditional U-Net, which leads to satisfying results, but it is possible that networks based on more state-of-the-art architectures such as nnUnet\cite{nn_unet} , transUnet\cite{transunet} , or visual transformer networks\cite{visual_transformer} , may outperform the original UNet structure used in this application.

For the future work, we will develop an improved ground truth for evaluation and investigate the performance of other architectures.

\acknowledgments
This work was supported in part by grants R01DC014037 and R01DC008408 from the NIDCD. This work is solely the responsibility of the authors and does not necessarily reflect the views of this institute.

% References
\bibliography{citation} % bibliography data in report.bib

\begin{thebibliography}{10}

\bibitem{rf}
RF, L. and JH., N., ``Preliminary results with image-guided cochlear implant
  insertion techniques,'' {\em Otology \& Neurotology}~{\bf 39}(7),  922--928
  (2018).

\bibitem{jh1}
Noble, J., Dawant, B., Warren, F., and Labadie, R., ``Automatic identification
  and 3d rendering of temporal bone anatomy,'' {\em Otology \&
  Neurotology}~{\bf 30}(4),  436--42 (2009).

\bibitem{gua}
Rohde, G.~K., Aldroubi, A., and Dawant, B.~M., ``The adaptive bases algorithm
  for intensity-based nonrigid image registration,'' {\em IEEE TRANSACTIONS ON
  MEDICAL IMAGING}~{\bf 22}(11),  1470--9 (2003).

\bibitem{andy}
Ding, A., Lu, A., Li, Z., Galaiya, D., Siewerdsen, J., Russell, T., and
  Creighton., F., ``Automated registration-based temporal bone computed
  tomography segmentation for applications in neurotologic surgery.,'' {\em
  Otolaryngol Head Neck Surg.}~{\bf 167}(1),  133–140 (2022).

\bibitem{wang2021}
Wang, J., Lv, Y., Wang, J., Ma, F., Du, Y., Fan, X., Wang, M., and Ke, J.,
  ``Fully automated segmentation in temporal bone ct with neural network: a
  preliminary assessment study,'' {\em BMC Medical Imaging}~{\bf 21},
  1471--2342 (2021).

\bibitem{unet}
Ronneberger, O., Fischer, P., and Brox, T., ``U-net: Convolutional networks for
  biomedical image segmentation,'' {\em CoRR}~{\bf abs/1505.04597} (2015).

\bibitem{cnn}
Krizhevsky, A., Sutskever, I., and Hinton, G.~E., ``Imagenet classification
  with deep convolutional neural networks,'' in [{\em Proceedings of the 25th
  International Conference on Neural Information Processing Systems - Volume
  1}{\nolinebreak\hspace{0.1em}]},  {\em NIPS'12},  1097–1105, Curran
  Associates Inc., Red Hook, NY, USA (2012).

\bibitem{bal}
Balakrishnan, G., Zhao, A., Sabuncu, M.~R., Guttag, J., and Dalca, A.~V.,
  ``Voxelmorph: a learning framework for deformable medical image
  registration,'' {\em IEEE transactions on medical imaging.}~{\bf 38}(8),
  1788--1800 (2013).

\bibitem{levelset}
Osher, S. and Fedkiw, R.~P., ``Level set methods: An overview and some recent
  results,'' {\em Journal of Computational Physics}~{\bf 169}(2),  463--502
  (2001).

\bibitem{nn_unet}
Isensee, F., Petersen, J., Klein, A., Zimmerer, D., Jaeger, P.~F., Kohl, S.,
  Wasserthal, J., K{\"{o}}hler, G., Norajitra, T., Wirkert, S.~J., and
  Maier{-}Hein, K.~H., ``nnu-net: Self-adapting framework for u-net-based
  medical image segmentation,'' {\em CoRR}~{\bf abs/1809.10486} (2018).

\bibitem{transunet}
Chen, J., Lu, Y., Yu, Q., Luo, X., Adeli, E., Wang, Y., Lu, L., Yuille, A.~L.,
  and Zhou, Y., ``Transunet: Transformers make strong encoders for medical
  image segmentation,'' {\em CoRR}~{\bf abs/2102.04306} (2021).

\bibitem{visual_transformer}
Dosovitskiy, A., Beyer, L., Kolesnikov, A., Weissenborn, D., Zhai, X.,
  Unterthiner, T., Dehghani, M., Minderer, M., Heigold, G., Gelly, S.,
  Uszkoreit, J., and Houlsby, N., ``An image is worth 16x16 words: Transformers
  for image recognition at scale,'' {\em CoRR}~{\bf abs/2010.11929} (2020).

\end{thebibliography}
\bibliographystyle{spiebib} % makes bibtex use spiebib.bst

\end{document}